\documentclass[%
 aip,
 sd,%
 amsmath,amssymb,
 reprint,%
]{revtex4-1}

\usepackage{graphicx}
\usepackage{dcolumn}
\usepackage{bm}
\usepackage{eurosym}
\usepackage{footmisc}
\usepackage{amssymb}
\usepackage{lipsum}
\usepackage{hyperref}
\usepackage{graphicx}
\usepackage{amsmath}
\usepackage{subfigure}

\DeclareGraphicsExtensions{.pdf,.eps,.png,.jpg,.mps}

\begin{document}

\preprint{AIP/123-QED}

\title{Stabilization of solitons under competing nonlinearities by external potentials}

\author{Krzysztof B. Zegadlo}
\email[]{zegadlo@if.pw.edu.pl}
\affiliation{ Faculty of Physics , Warsaw University of Technology,
Warsaw, ul. Koszykowa 75, PL–00–662 Warszawa, Poland}
\author{Tomasz Wasak} 
\affiliation{Faculty of Physics, University of Warsaw, ul. Ho˙za 69, PL–00–681 Warszawa, Poland}
\author{Boris A. Malomed}
\affiliation{Department of Physical Electronics, School of Electrical Engineering,
Faculty of Engineering, Tel Aviv University, Tel Aviv 69978, Israel}
\author{Miroslaw A. Karpierz}
\affiliation{ Faculty of Physics , Warsaw University of Technology,
Warsaw, ul. Koszykowa 75, PL–00–662 Warszawa, Poland}
\author{Marek Trippenbach}
\affiliation{Faculty of Physics, University of Warsaw, ul. Ho˙za 69, PL–00–681 Warszawa, Poland}

\date{\today}

\begin{abstract}
We report results of the analysis for families of one-dimensional (1D)
trapped solitons, created by competing self-focusing (SF) quintic and
self-defocusing (SDF) cubic nonlinear terms. Two trapping potentials are
considered, the harmonic-oscillator (HO) and delta-functional ones. The
models apply to optical solitons in colloidal waveguides and other photonic media, and to matter-wave solitons in Bose-Einstein condensates (BEC) loaded into a quasi-1D trap. For the HO potential, the results are obtained in an approximate form, using the variational and Thomas-Fermi approximations (VA and TFA), and in a full numerical form, including the ground state and the first antisymmetric excited one. For the delta-functional attractive potential, the results are produced in a fully analytical form, and verified by means of numerical methods. Both exponentially localized solitons and weakly localized trapped modes are found for the delta-functional potential. The most essential conclusions concern the applicability of \emph{competing} Vakhitov-Kolokolov (VK) and anti-VK criteria to the identification of the stability of solitons created under the action of the competing SF and SDF terms.
\end{abstract}

\pacs{Valid PACS appear here}
\keywords{Suggested keywords}
\maketitle

\begin{quotation}
Solitons, i.e., nondiffracting beams propagating in nonlinear media, are formed thanks to an interplay between diffraction, which tends to spread the beam, and the self-induced change of the nonlinear refractive index of the medium. Optical solitons are important for their capability to beat diffraction, and their potential for engineering a variety of reconfigurable optical structures including (and not limited to) couplers, deflectors, and logic gates. A classical example of the setting supporting solitons is provided by Kerr media, featuring the third-order self-focusing nonlinearity. However, more complex optical media (for instance, colloids filled by metallic nanoparticles) often exhibit competing nonlinearities of different orders, most typically self-focusing cubic and self-defocusing quintic. In the latter case, stable optical solitons also propagate through the medium, differing from the Kerr solitons by a broader shape. More challenging is the opposite case of the competition between the lower-order (cubic) self-defocusing and higher-order (quintic) self-focusing nonlinearities. In the uniform medium, this combination may generate only strongly unstable solitons. In this work we demonstrate that the latter situation can be remedied by an effective trapping potential. In term of optical media, the potential represents a waveguiding structure, and, by itself, it is a linear ingredient of the system. We predict that trapping potentials, both tight and loose, readily stabilize vast families of solitons, which are completely unstable in the uniform medium.
\end{quotation}

\section{Introduction}

\label{sec:introduction}

Competition between self-focusing (SF) and self-defocusing (SDF)\
nonlinearities occurs in various physical media, playing an 
important role in the creation of self-trapped modes (in particular,
solitons). A well-known example is the competition between
quadratic (second-harmonic-generating, alias $\chi ^{(2)}$) and SDF cubic
nonlinear interactions in optics, in the case when the proper choice of the
mismatch constant makes the $\chi ^{(2)}$ interaction effectively
self-focusing \cite{chi2_1,chi2_2,chi2_3,chi2_4,chi2_5,chi2_6,chi2_7,chi2-review,chi2-review2}.

A large number of publications have addressed systems featuring the competition
between SF cubic and SDF quintic terms. Such combinations of nonlinear terms
frequently occur in optics, including liquid waveguides \cite%
{liquids,liquids2,liquids3}, special kinds of glasses \cite%
{liquids,quintic-experiment,quintic-experiment2,quintic-experiment3} and
ferroelectric films \cite{ferroelectric}. Especially flexible are colloidal
media formed by metallic nanoparticles, in which the cubic-quintic (CQ)
optical nonlinearity can be adjusted within broad limits by selecting the
size of the particles and the colloidal filling factor \cite%
{colloid,colloid2,colloid3,colloid4}.

The SF-SDF CQ nonlinearity has a great potential for the creation of stable
multidimensional solitons, including two- \cite%
{Manolo,Manolo2,Manolo3,Pramana} and three- \cite{nine-authors,nine-authors2}
dimensional (2D and 3D) solitary vortices, as reviewed in Ref. \cite{review}
and recently demonstrated experimentally in the 2D setting, in the
colloidal waveguide, in Ref. \cite{liquids3}. Indeed, the cubic-only SF
nonlinearity cannot create stable multidimensional solitons, as the
corresponding 2D self-trapped modes (alias \textit{Townes' solitons}) and 3D
solitons are subject to instabilities related to the critical collapse in
2D [recently, it was demonstrated that stable 2D composite 
(half-fundamental - half-vortical) solitons can be 
created in a two-component system combining the cubic self-attraction 
and linear mixing of the components through first-order 
spatial derivatives, which represent the spin-orbit coupling\cite{HS1}], and supercritical collapse in 3D \cite{Berge,Berge2}(spatial inhomogeneity, in the form of a finite jump of the SF Kerr coefficient between an inner circle and the surrounding area, may stabilize fundamental solitons \cite{HS2}). The additional SDF quintic term arrests the collapse, imposing soliton stability \cite{Manolo,Manolo2,Manolo3,Pramana,review,nine-authors,nine-authors2}.

The study of the competing CQ nonlinearities is also relevant in 1D
settings. A remarkable fact is that, although the 1D nonlinear Schr\"{o}%
dinger (NLS) equation with the combined CQ nonlinearity is not integrable,
it admits well-known exact solutions for the full soliton family \cite%
{Pushkarov,Pushkarov2}. Furthermore, these solitons are stable not only in
the case of the competition between the SF cubic and SDF quintic terms, but
also when both terms have the SF sign \cite{Pelin}. The latter fact is
surprising, because the 1D solitons created by the quintic-only SF term are,
as a matter of fact, a 1D version of the Townes' solitons \cite%
{Salerno,Salerno2,Salerno3}, and, accordingly, are unstable against the 1D
variety of the critical collapse. When both cubic and quintic SF terms are
present, the 1D solitons may still be pushed into the collapse by strong
perturbations (sudden compression), but, somewhat counter-intuitively, they
are stable against small perturbations.

In the case of the combination of the SDF cubic and SF
quintic terms, the entire family of 1D solitons is completely unstable. We stress that this
combination of the competing nonlinearities may be physically relevant.
In addition to the optical examples mentioned above, it appears naturally as a result of the reduction of the 3D
Gross-Pitaevskii (GP) equation to the 1D form for cigar-shaped traps filled
with atomic Bose-Einstein condensates (BECs) \cite%
{1D-BEC,1D-BEC2,1D-BEC3,1D-BEC4,1D-BEC5,1D-BEC6}. In the latter case, the
sign of the cubic terms is determined by the sign of the scattering length
in the BEC, and typically corresponds to the SDF, while the quintic term is
generated by the tight confinement of the condensate in the transverse
plane, and always corresponds to the SF.

Thus, search for physically realistic settings that would allow the
stabilization of 1D solitons in the NLS equation with the SDF-SF combination
of the cubic and quintic terms is an interesting problem, which is the subject
of the present work. It is well known that 2D Townes solitons and solitary
vortices, created by the SF cubic nonlinearity, can be readily stabilized by
an harmonic-oscillator (OH) trapping potential \cite%
{2D-HO-stabilization,2D-HO-stabilization2,2D-HO-stabilization3,2D-HO-stabilization4,2D-HO-stabilization5}%
, or by periodic potentials provided by optical lattices \cite%
{2D-OL-stabilization,2D-OL-stabilization2}. In this work, we consider the
stabilization of 1D solitons provided by the HO potential, and also by the
delta-functional attractive potential. In the optical waveguide, effective
traps may be induced by a transverse profile of the refractive index
\cite{Kivshar-Agrawal}, while in the BEC setup traps are created by means of
properly tailored magnetic and/or optical fields \cite{Pitaevskii,Randy}.
In these contexts, both broad HO traps and narrow ones, which may be
approximated by the Dirac's delta-function, are in principle relevant settings.

The NLS equation with the SDF-SF CQ nonlinearity and a trapping potential
offers a possibility to explore peculiarities of the stabilization of 1D
solitons under the action of the competing nonlinear terms, which is
especially interesting as concerns the central issue of this topic, i.e.,
identification of the stability of the so produced solitons. Indeed, it is
commonly known that the necessary stability criterion for families of
solitons supported by the SF nonlinearity is provided by the Vakhitov-
Kolokolov (VK) criterion \cite{VK,Berge,Berge2}, which is simply formulated
in terms of the dependence between the soliton's norm (total power), $N$,
and its propagation constant, $k$: $dN/dk>0$. In some cases, this stability
criterion may actually be a sufficient one, too. On the other hand, for
solitons supported by the interplay of SDF nonlinearity and trapping
potentials, a necessary stability condition is provided by the \textit{%
anti-VK criterion}, $dN/dk<0$ \cite{HS}. Therefore, a natural question,
which we address in the present work, is what criterion determines the
stability of solitons in the case of the competition between the SDF cubic
and SF quintic terms. In the opposite case of the combination of the SF
cubic and SDF quintic nonlinearities, the former one dominates the formation
of 1D solitons, their entire family being stable in accordance with the VK
criterion.

The rest of the paper is organized as follows. The model is formulated in
Section II. Findings for the soliton trapped in the HO potential are
collected in Section III, which includes both approximate analytical
results, obtained by means of the variational and Thomas-Fermi approximations
(VA and TFA correspondingly), and systematically generated numerical results for the
existence and stability of the solitons. A significant conclusion is that the
usual VK criterion works, as the necessary and sufficient one, for \textit{%
forward-going} branches of $N(k)$ curves of soliton families, in either case
of the positive and negative local slope ($dN/dk\gtrless 0$), while the
anti-VK criterion works equally well, but for \textit{backward-going}
branches. Both these types of branches are produced by the analytical
and numerical results under the action of the HO trap. Findings for the
delta-functional potential are reported in Section IV, including both
exponentially localized solitons and weakly localized but normalizable
trapped modes. In fact, they are obtained in a fully analytical form, which
is verified by dint of the numerical analysis, and it is again demonstrated
that the usual VK criterion is correct for all the solitons in the case of
forward-going branches, while the anti-VK is proper for backward-going ones.
The paper is concluded by Section V.

\section{The model}

\label{sec:model} In the free space, the 1D NLS equation with the SDF cubic
and SF quintic terms is
\begin{equation}
i\frac{\partial u}{\partial z}+\frac{1}{2}\frac{\partial ^{2}u}{\partial
x^{2}}-g|u|^{2}u+|u|^{4}u=0,  \label{NLSE}
\end{equation}%
where $g\geq 0$ is the strength of the cubic nonlinearity. Here, Eq. (\ref%
{NLSE}) is written in the form of the paraxial equation for the spatial
evolution of the optical beam in a nonlinear planar waveguide, $z$ and $x$
being the longitudinal and transverse coordinates, respectively \cite%
{Kivshar-Agrawal}. In physical units, the solitons considered below may have
the transverse width $\lesssim 50$ $\mathrm{\mu }$m, while the
experimentally relevant transmission distance may be a few cm. If Eq. (\ref%
{NLSE}) is considered as the reduced GP equation for the BEC, the solitons
may be composed of several thousands of atoms, with the characteristic size $%
\lesssim 100$ $\mathrm{\mu }$m \cite{Randy}.

Equation (\ref{NLSE}) has a family of exact soliton solutions with
propagation constant $k\geq 0$, which can be easily obtained from the
well-known solitons for the opposite case of the SF-SDF CQ nonlinearity \cite%
{Pushkarov} by means of the analytical continuation:
\begin{equation}
u\left( x,z\right) =e^{ikz}\sqrt{\frac{2\sqrt{3}k}{\sqrt{4k+3g^{2}/4}\cosh
\left( {2\sqrt{2k}x}\right) -(g/2)\sqrt{3}}}  \label{solution}
\end{equation}%
(in the case of the GP equation, $-k$ is the chemical potential of the BEC).
The norm of the soliton (\ref{solution}), which represents the total power of
the spatial optical solitons, or the total number of atoms in matter-wave
solitons in BEC, is
\begin{widetext}
\begin{equation}
N(k)\equiv \int_{-\infty }^{+\infty }|u(x)|^{2}dx=\sqrt{6}\arctan \left( {%
\frac{2\sqrt{k}}{\sqrt{4k+3g^{2}/4}-(g/2)\sqrt{3}}}\right) .  \label{N}
\end{equation}%
\end{widetext}
This soliton family is completely unstable. Notice that $N(k)$, as given by Eq. (%
\ref{N}) yields $dN/dk<0$, hence it does not satisfy the VK criterion \cite%
{VK}.

As said above, solitons can be stabilized by means of an external
potential, $W(x)$, which added to Eq. (\ref{NLSE}) gives:%
\begin{equation}
i\frac{\partial u}{\partial z}+\frac{1}{2}\frac{\partial ^{2}u}{\partial
x^{2}}-g|u|^{2}u+|u|^{4}u=W(x)u.  \label{U}
\end{equation}%
The stationary version of Eq. (\ref{U}), corresponding to $u\left(
x,z\right) =\exp \left( ikz\right) U(x)$, is
\begin{equation}
-kU+\frac{1}{2}U^{\prime \prime }-gU^{3}+U^{5}=W(x)U.
\label{stationary equation}
\end{equation}%
In the optical waveguide, the potential represents an inhomogeneous profile
of the refractive index, while in the BEC trap it may be induced by an external laser beam. In either type of the physical situation imposed potential may be either broad
(width measured by dozens of $\mathrm{\mu }$m) or narrow o(squeezed to a few $\mathrm{\mu }$m). Here we consider two
basic types of the trapping potential, harmonic trap,
\begin{equation}
W(x)=({1/2})\Omega ^{2}x^{2},  \label{HO}
\end{equation}%
and the Dirac's delta-function,
\begin{equation}
W(x)=-\epsilon \delta (x)  \label{delta}
\end{equation}%
with $\epsilon >0$. By the means of rescaling, we fix $\Omega \equiv 0.5$ in
Eq. (\ref{HO}), and $\epsilon =0.5$ in Eq. (\ref{delta}).

The delta-functional attractive potential (\ref{delta}) implies that the $x$%
-derivative of the wave field is a subject to a jump condition at $x=0$:%
\begin{equation}
u_{x}\left( x=+0\right) -u_{x}\left( x=-0\right) =-2\epsilon u(x=0),
\label{jump}
\end{equation}%
while the field itself is continuous at this point. In fact, a full family of
solitons pinned to the ideal delta-functional potential embedded into the
medium with the CQ nonlinearity can be found in an exact analytical form, as
shown below in Section IV.

\section{The harmonic-oscillator potential}

\label{sec:harmonic}

In this section we present results obtained for Eq. (\ref{stationary equation}) with the HO
potential (\ref{HO}). We first use analytical approximations to gain a first
insight as to where one may expect solitons, and then we concentrate on
numerical analysis in these regions.

\subsection{The variational approximation (VA)}

\label{subsec:VA}

The stationary NLSE equation is now given by
\begin{equation}
-kU+\frac{1}{2}U^{\prime \prime }-gU^{3}+U^{5}=\frac{1}{2}\Omega ^{2}x^{2}U,
\label{stationary NLSE}
\end{equation}%
where $k$ is the propagation constant, the prime stands for $d/dx$, and the
HO potential is introduced as per Eq. (\ref{HO}). The starting point of the
VA is that Eq. (\ref{stationary NLSE}) can be derived from the Lagrangian
density,
\begin{equation}
2\mathcal{L}=\frac{1}{2}\left( U^{\prime }\right) ^{2}+kU^{2}+\frac{1}{2}%
gU^{4}-\frac{1}{3}U^{6}+\frac{1}{2}\Omega ^{2}x^{2}U^{2}.
\label{Lagrangian density}
\end{equation}

\begin{widetext}

\begin{figure}[th]
\centering
\includegraphics[scale=0.22]{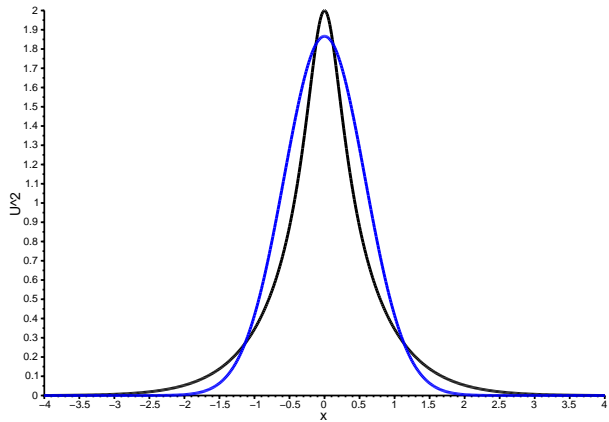}
\caption{The wave function of the fundamental mode trapped in the HO
potential (\protect\ref{HO}) with $\Omega =0.5$, as obtained in the
numerical form, by means of the shooting method (the black, taller profile),
and as predicted analytically by the VA (the blue, lower profile) for $g=2$,
$k=1.309$, $N=2.69$. The norm is calculated for the numerical solution.}
\label{ITM+VA2}
\end{figure}

\end{widetext}

We approximate the fundamental trapped mode by the usual Gaussian \textit{%
ansatz} \cite{ansatz,ansatz2},
\begin{equation}
u(x)=A\exp \left( -x^{2}/w^{2}\right) ,  \label{VA solution}
\end{equation}%
with amplitude $A$, width $w$, and norm%
\begin{equation}
N_{\mathrm{VA}}=\sqrt{\pi /2}A^{2}w.  \label{NVA}
\end{equation}%
The substitution of ansatz (\ref{VA solution}) into Eq. (\ref{Lagrangian
density}) and integration over $x$ yields the corresponding effective
Lagrangian, $L=\int_{-\infty }^{+\infty }\mathcal{L}(x)dx$,
\begin{widetext}
\begin{equation}
\frac{2L}{\sqrt{\pi }}=\frac{1}{2}\left( \frac{2}{w^{4}}+\frac{1}{2}\Omega
^{2}\right) A^{2}\left( \frac{w}{\sqrt{2}}\right) ^{3}+\frac{w}{\sqrt{2}}%
kA^{2}+\frac{1}{4}gA^{4}w-\frac{1}{3}A^{6}\frac{w}{\sqrt{6}},
\label{Lagrangian}
\end{equation}%
\end{widetext}
which gives rise to the respective Euler-Lagrange equations, $\partial
L/\partial A^{2}=\partial L/\partial w=0$, i.e.,
\begin{equation}
\frac{1}{2}\left( \frac{2}{w^{4}}+\frac{1}{2}\Omega ^{2}\right) \left( \frac{%
w}{\sqrt{2}}\right) ^{3}+\frac{kw}{\sqrt{2}}+\frac{1}{2}gA^{2}w-A^{4}\frac{w%
}{\sqrt{6}}=0,  \label{dL/dA2}
\end{equation}%
\begin{equation}
-\frac{1}{w^{2}\sqrt{2}}+\frac{3}{4}\Omega ^{2}\frac{w^{2}}{(\sqrt{2})^{3}}+%
\frac{k}{\sqrt{2}}+\frac{1}{4}gA^{2}-\frac{1}{3}\frac{A^{4}}{\sqrt{6}}=0.
\label{dL/dw}
\end{equation}%
Equation (\ref{dL/dw}) produces two solutions for the squared amplitude,
\begin{equation}
A_{\pm }^{2}=\sqrt{\frac{3}{8}}g\pm \sqrt{\frac{3}{8}g^{2}+\frac{\sqrt{3}}{4}%
\left( \frac{2}{w^{2}}+\frac{1}{2}\Omega ^{2}w^{2}+4k\right) }.  \label{A^2}
\end{equation}%
Solution $A_{+}^{2}$ is physical (positive) under condition $k\geq -\frac{%
\sqrt{3}}{8}g^{2}-\frac{1}{2}w^{-2}-\frac{1}{8}\Omega ^{2}w^{2}$, while $%
A_{-}^{2}$ is relevant only for $-\frac{\sqrt{3}}{8}g^{2}-\frac{1}{2}w^{-2}-%
\frac{1}{8}\Omega ^{2}w^{2}\leq k<-\frac{1}{2}w^{-2}-\frac{1}{8}\Omega
^{2}w^{2}$. In Fig. (\ref{ITM+VA2}) a typical fundamental-mode profile
produced by the VA is compared to its numerical counterpart.

\subsection{The Thomas-Fermi approximation (TFA)}

For negative values of $k$ in Eq. (\ref{stationary NLSE}), another
analytical approach may be applied, in the form of the TFA, which neglects
the diffraction term, $d^{2}U/dx^{2}$, in the equation. This approximation,
which is relevant for confined modes when the SDF nonlinear term is the
dominant one \cite{Borovkova,Borovkova2}, yields the stationary solution in the form of
\begin{equation}
U_{\mathrm{TFA}}^{2}(x)=\left\{
\begin{array}{c}
\frac{g}{2}-\sqrt{\frac{g^{2}}{4}-\left( |k|-\frac{1}{2}\Omega
^{2}x^{2}\right) },~\mathrm{at}~~x^{2}<2|k|/\Omega ^{2}, \\
0,~\mathrm{at}~~x^{2}>2|k|/\Omega ^{2}.%
\end{array}%
\right.   \label{TF}
\end{equation}%
As seen from Eq. (\ref{TF}), the TFA solution is a physically relevant
provided that the wavenumber satisfies condition $0<-k<g^{2}/4$. There exists also
a solution with the positive sign in front of the square root in Eq. (%
\ref{TF}), but it is irrelevant, as it does not vanish at $|x|\rightarrow
\infty $. The comparison of typical solution profiles predicted by the TFA
with their numerical counterparts is presented in Fig. \ref{ITM+TF for g=2,4}%
. It can be concluded that the TFA predicts the solutions in a qualitatively
correct form only when the cubic SDF term dominates in Eq. (\ref{stationary
NLSE}), which occurs at $g\gtrsim 2$.

\begin{widetext}

\begin{figure}[th]
\centering\subfigure[]{\includegraphics[scale=0.2]{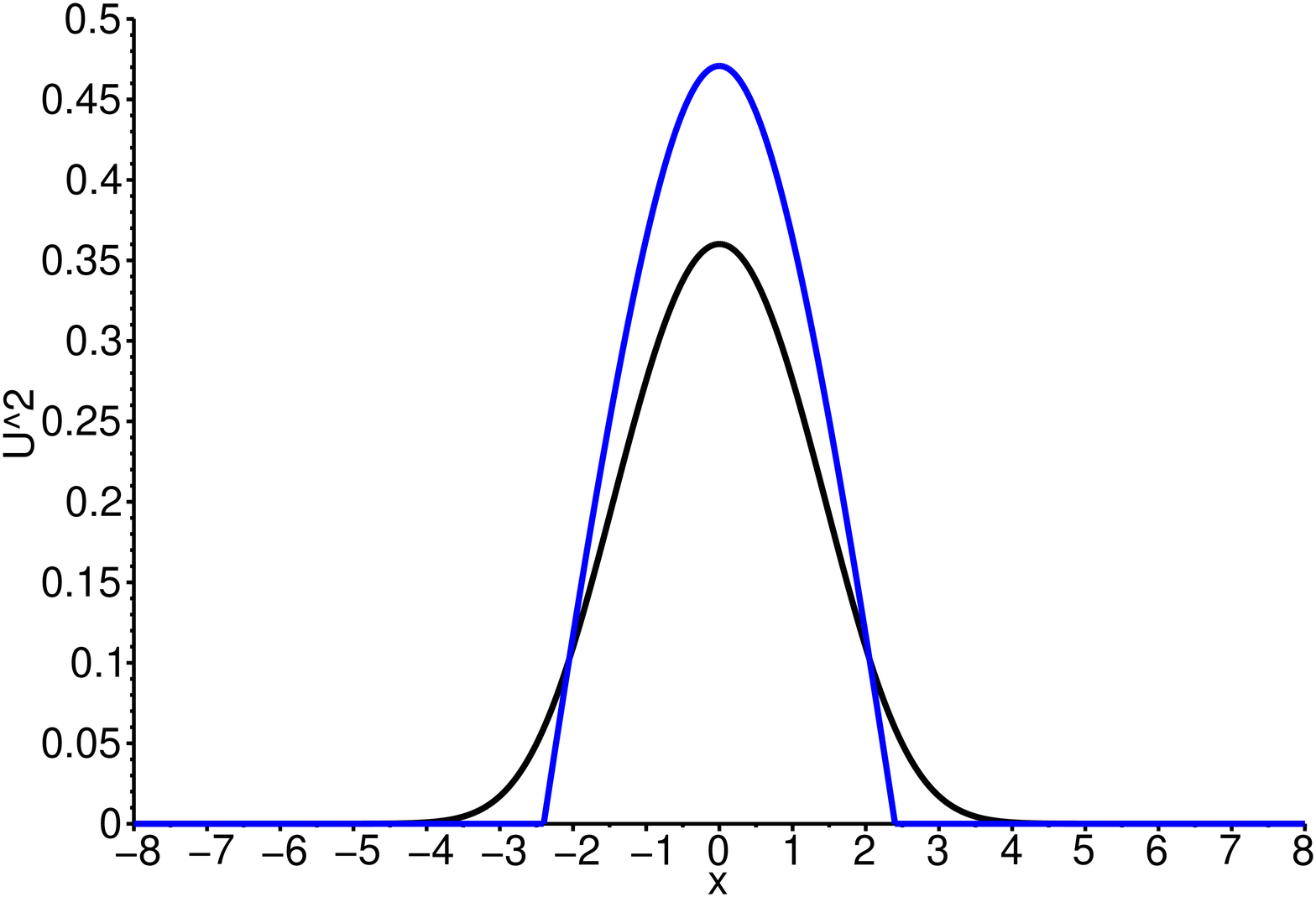}}%
\subfigure[]{\includegraphics[scale=0.2]{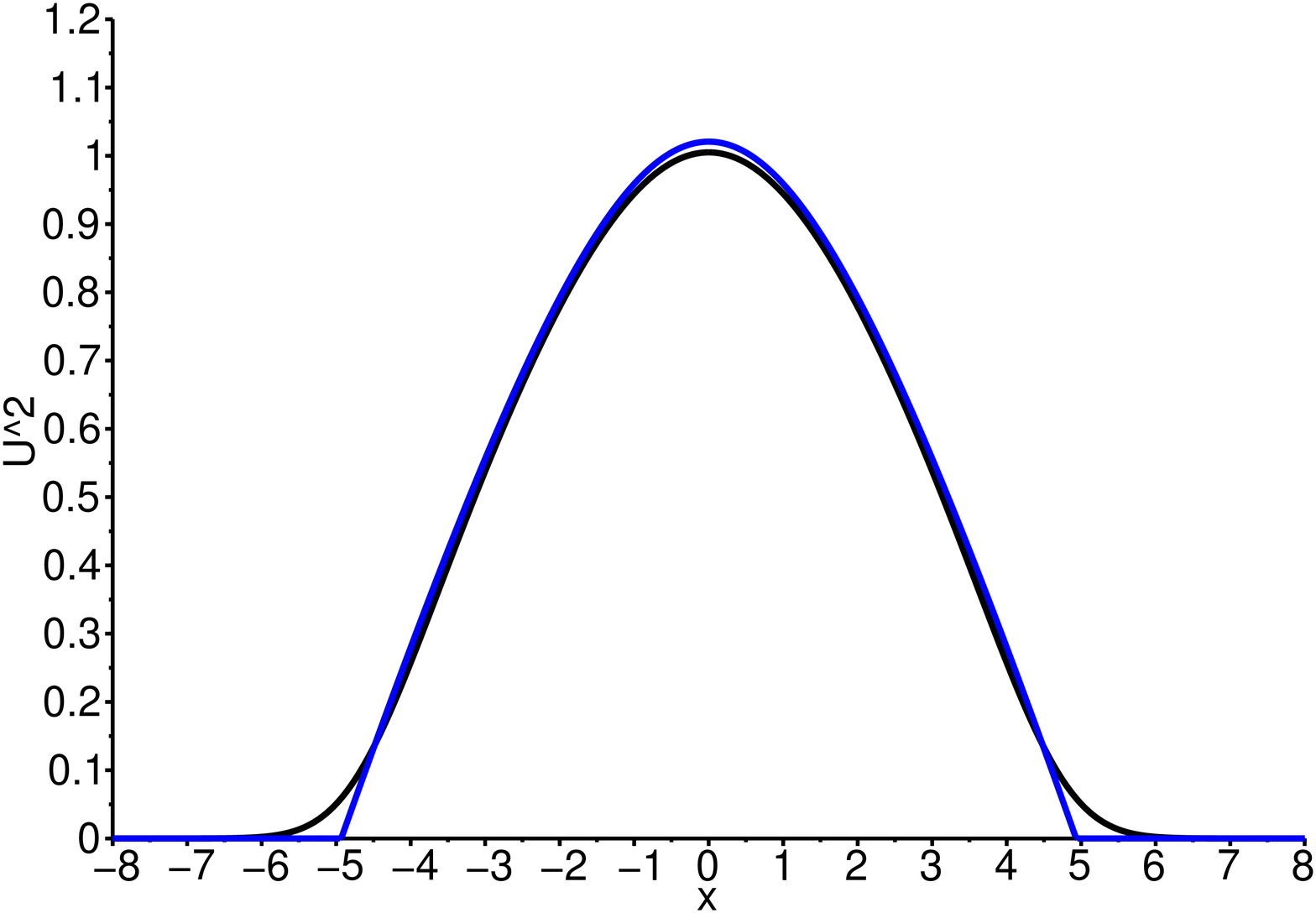}}
\caption{Examples of the fundamental mode trapped in the HO potential, as
obtained in the numerical form (the black, lower profiles), and by means of
the TFA (the blue profiles with truncated tails), for $g=2,k=-0.72,N=1.17$
(a) and $g=4,k=-3.041,N=6.15$ (b).}
\label{ITM+TF for g=2,4}
\end{figure}

\end{widetext}

\subsection{Numerical results}

\label{sec:numerical}

\subsubsection{Fundamental solutions}

\label{subsubsec:symmetric solutions}

\begin{widetext}

\begin{figure}[th]
\centering
\includegraphics[scale=0.33]{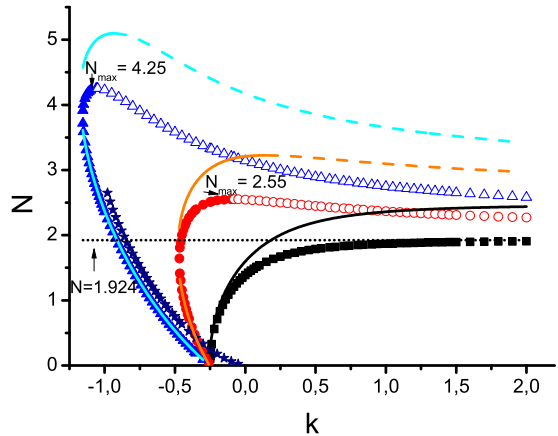}
\caption{The norm of the fundamental modes, trapped in the HO potential (%
\protect\ref{HO}) versus $k$ for different values of $g$. Here and in Fig.
\protect\ref{N-antisymmetric} below, full and empty marks designate,
respectively, stable and unstable solutions found numerically: black (squares) for $g=0
$, red (circles) for $g=1$, and blue (triangles) for $g=2$. Solid and dashed lines depict
VK-stable and unstable states predicted by the VA: black (the lowest line in the graph) for $g=0$, orange (the middle one)
for $g=1$ and cyan (the highest position in the graph) for $g=2$. Results obtained for $g=2$ by the means of TFA are depicted by navy-blue stars. The black dotted horizontal line here and in Fig. \protect\ref{N-antisymmetric} corresponds to the limit value of $N(k=\infty)=(\sqrt{6}/4)\pi\approx 1.924$, see Eq. (\ref{N}), which does not depend on $g$. Higher order modes are not included here.}
\label{N(k)-symmetric}
\end{figure}

\end{widetext}

Numerical stationary solutions for HO potential were
found by means of the imaginary-time-integration \cite{IT} and shooting \cite%
{shooting} methods. The results are collected in Fig. \ref{N(k)-symmetric},
where the corresponding dependences $N(k)$ are displayed for different
values of the cubic-SDF coefficient, $g$, and compared to the results
produced by the VA and TFA. All the branches stem, at $N=0$, from the point which corresponds to the fundamental mode of the HO potential in linear quantum mechanics.

The maximum value of dependence $N(k)$ increases with the growth of $g$. The turning point, where the two branches of the VA solution meet, can be
estimated from Fig. \ref{N(k)-symmetric}. For instance, it is $k\approx -1$
at $g=2$. Only a small part of the top (blue) solution branch in Fig. \ref%
{N(k)-symmetric} satisfies the Vakhitov-Kolokolov (VK) criterion, $dN/dk>0$,
which, as said above, is a necessary condition for the stability of
localized modes supported by the SF nonlinearity \cite{VK}. On the other
hand, the necessary stability condition for localized modes dominated by the
SDF nonlinearity may switch from the VK form to the \textit{anti-VK} one, $%
dN/dk<0$ \cite{HS}. Because the present model features the competition of
the quintic SF and cubic SDF terms, it is not obvious which one plays the
dominant role at different values of $k$ and $N$, hence a numerical stability analysis is necessary.

To explore the stability of the solutions, eigenfrequencies of small
perturbations around the stationary solutions have been computed, using the
standard linearization procedure \cite{Yang}. For stable solutions,
imaginary parts of these frequencies are zero (or negligibly small, in terms
of the numerical computation). All instabilities detected by this method
feature pure-imaginary eigenvalues (i.e., the respective instability mode is
expected to grow exponentially without oscillations). The latter finding
suggests that the VA and anti-VK criteria may be sufficient for the
stability analysis, because what they cannot detect, are complex unstable
eigenvalues \cite{VK,Berge}, which do not exists in the present setting
anyway. Eventually, the full results for the (in)stability eigenvalues
demonstrate that the VA criterion \emph{correctly} predicts the
(in)stability for the \textit{forward-going} segments of the $N(k)$
branches, with either sign of the slope, $dN/dk\gtrless 0$, see Fig. \ref%
{N(k)-symmetric}, while the anti-VK criterion is also \emph{correct}, but for
the \textit{backward-going} branches. In fact, the latter ones always have $%
dN/dk<0$, being completely stable, accordingly. Our findings imply a
conclusion that may be relevant for other models too, \textit{viz}., that
the SF and SDF nonlinearities determine the stability, severally, for the
forward- and backward-going segments of families of trapped states.

The stability analysis presented above has been corroborated by direct
simulations of the perturbed evolution of the corresponding trapped states.
The simulations were run by dint of the finite-difference algorithm. In
particular, the solutions predicted to be unstable indeed blow up  in the
course of the evolution (i.e., exhibit the collapse driven by the SF quintic
term \cite{Berge}; not shown here in detail).

\subsubsection{Antisymmetric trapped states}

\label{subsubsec:antisymmetric solutions}

\begin{widetext}

\begin{figure}[th]
\centering
\includegraphics[scale=0.2]{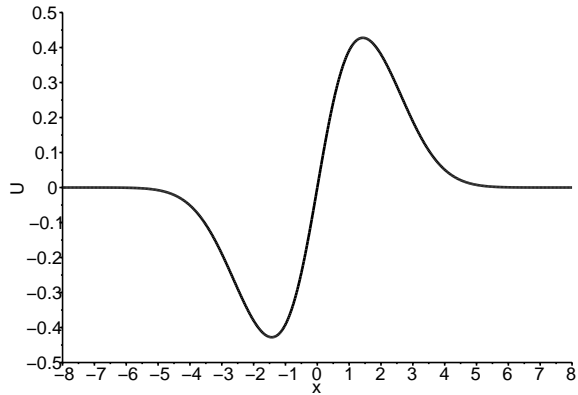}
\caption{The profile of the antisymmetric excited state, trapped in the HO
potential for $g=2$, $k=-1$, and $N=0.673$.}
\label{antisymmetric profile}
\end{figure}

\end{widetext}

\begin{widetext}

\begin{figure}[th]
\centering
\includegraphics[scale=0.35]{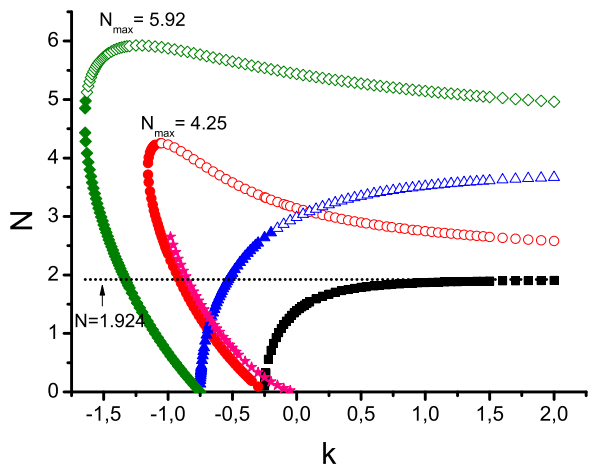}
\caption{Numerically obtained $N(k)$ dependences for symmetric modes trapped in the HO potential,
at $g=0$ (black, squares) and $g=2$ (red, circles), and for antisymmetric ones at $g=0$ (blue, triangles)
and $g=2$ (green, turned squares). The TFA solutions for $g=2$ are designated by pink stars. The third and higher order modes are not included here.}
\label{N-antisymmetric}
\end{figure}

\end{widetext}

A family of the lowest excited states with an antisymmetric profile,
has been found and
investigated for the HO trap. Results are presented in Fig. \ref{antisymmetric profile}. Numerical solutions were constructed by
means of the shooting method, and then their stability was tested
through the computation of perturbation eigenvalues. In Fig.
\ref{N-antisymmetric} for the sake of comparison, branches
of the antisymmetric states are shown along with their ground-state
(symmetric) counterparts. Similar to the situation shown above, for the
ground-state branches in Fig. \ref{N(k)-symmetric}, all the curves
representing the antisymmetric states stem, at $N=0$, from the point
corresponding to the first excited state of the HO potential in
linear quantum mechanics.

\begin{widetext}

\begin{figure}[th]
\centering
\subfigure[]{\includegraphics[scale=0.19]{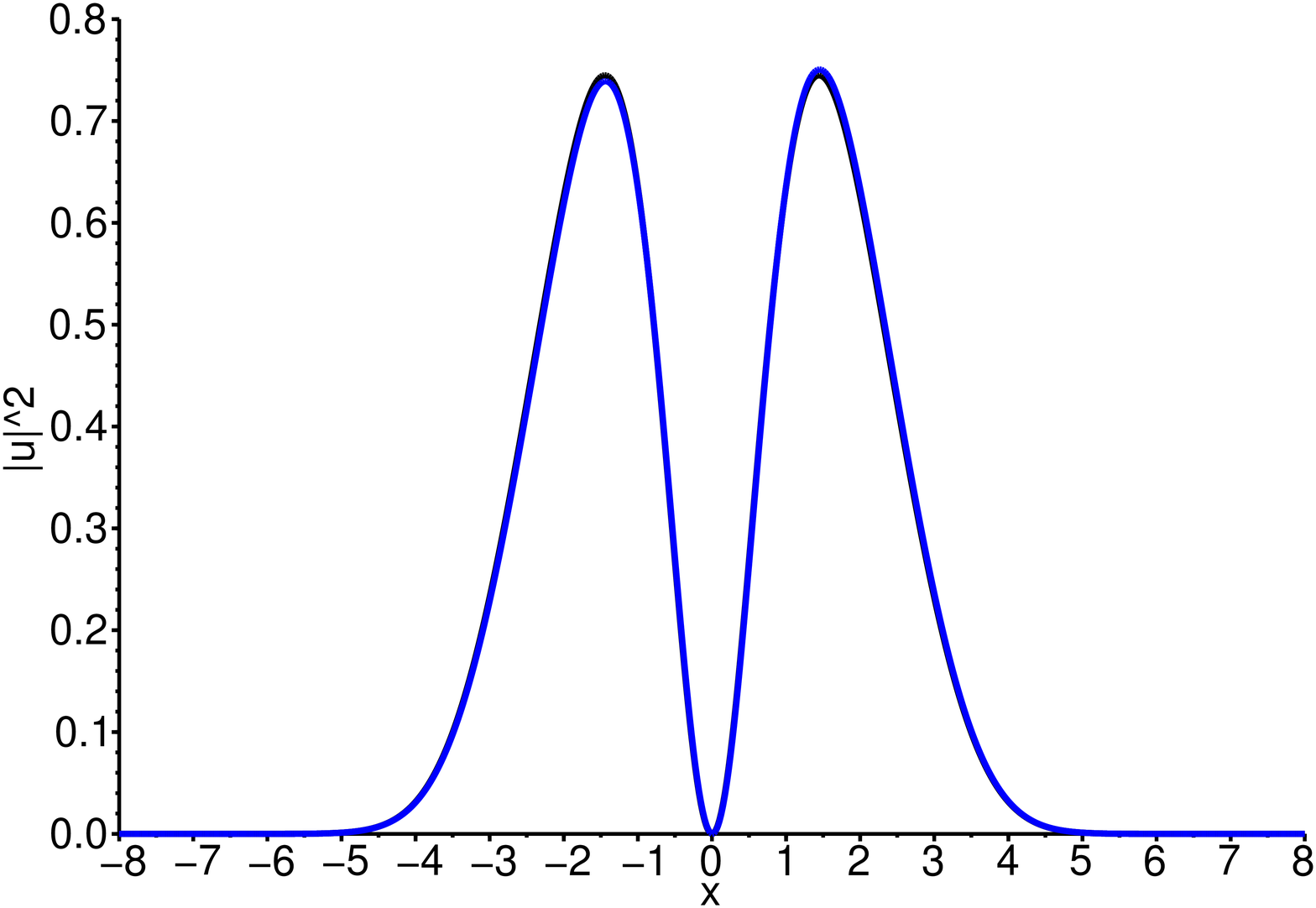}} %
\subfigure[]{\includegraphics[scale=0.19]{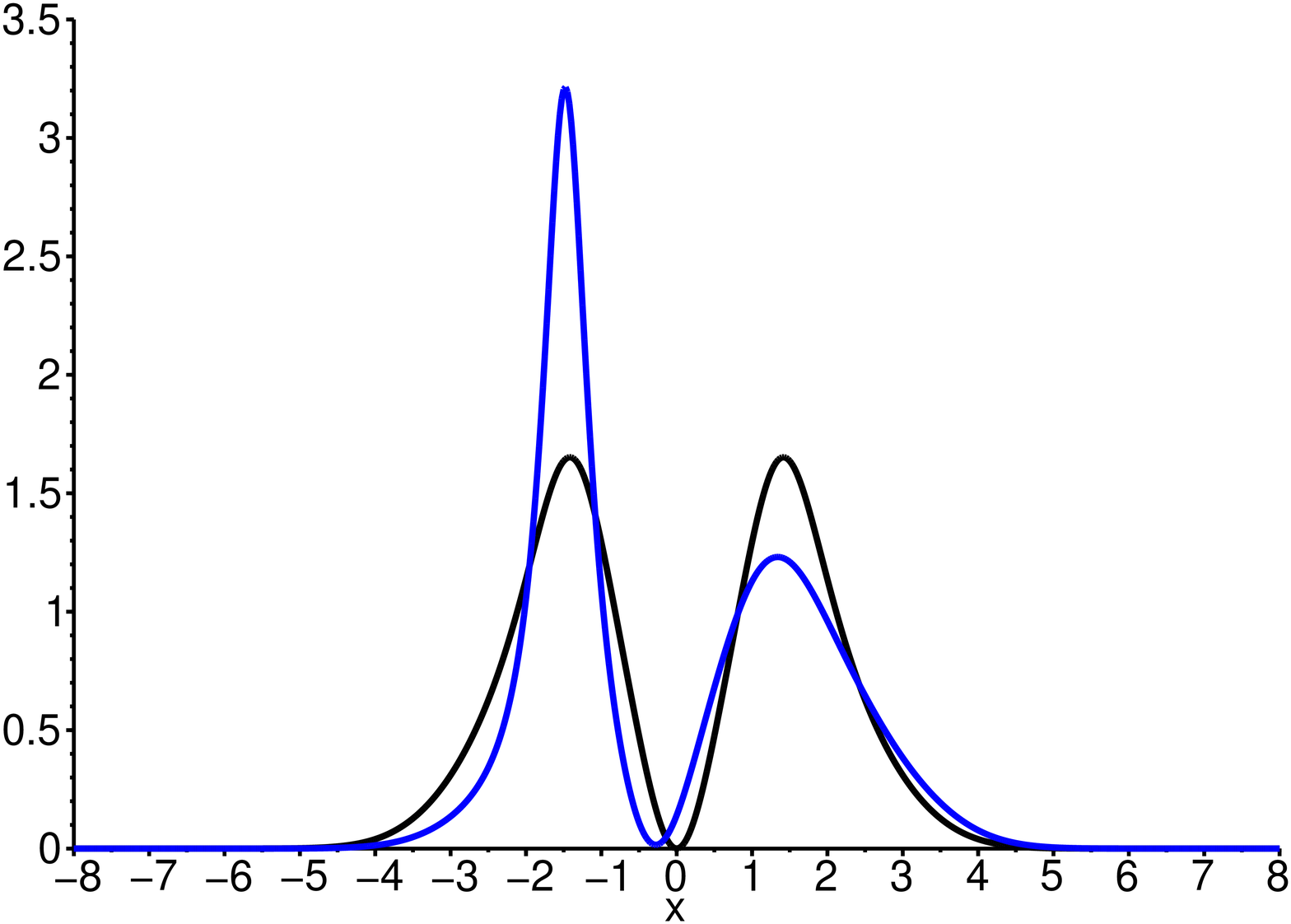}}
\caption{Stable (a) and unstable (b) evolution of an antisymmetric state
trapped in the HO potential, for $g=2$, $k=-1.53$, $N=3.067$, and $g=2$, $%
k=-1.53$, $N=5.657$, respectively (both solutions have the same $k$ but
belong, severally, to different branches: \textit{backward-going} and
\textit{forward-going}). The initial slightly perturbed stationary state and
the final one are shown by the black and blue curves, respectively.}
\label{antisymmetric-evolution-HO}
\end{figure}

\end{widetext}

The largest value of dependence $N(k)$ increases with the mode's number, 
i.e., the maximum of the green branch (antisymmetric mode) exceeds the 
maximum of the red one (symmetric mode). Furthermore, the plots presented in Fig. \ref{N-antisymmetric} demonstrate additional
instability of the antisymmetric modes, in comparison with their fundamental
symmetric counterparts (which is not surprising, as excited states in
nonlinear systems are usually more prone to instability \cite{Borovkova,Borovkova2}).
This instability concerns the forward-going branches, that satisfy the VK
criterion. The respective perturbation eigenvalues are complex, on the
contrary to the pure imaginary ones that could account for the instability
of the symmetric states, as mentioned above, hence they definitely cannot be
detected by the VK criterion, which is, generally, less relevant for excited
states, in comparison with the ground state. The predicted instability was
verified by direct simulations. It was found that the instability destroys
the antisymmetry of the stationary mode, and one of the two resulting peaks
develops the collapse, see Fig. \ref{antisymmetric-evolution-HO}.

\begin{figure}[th]
\centering
\includegraphics[scale=0.36]{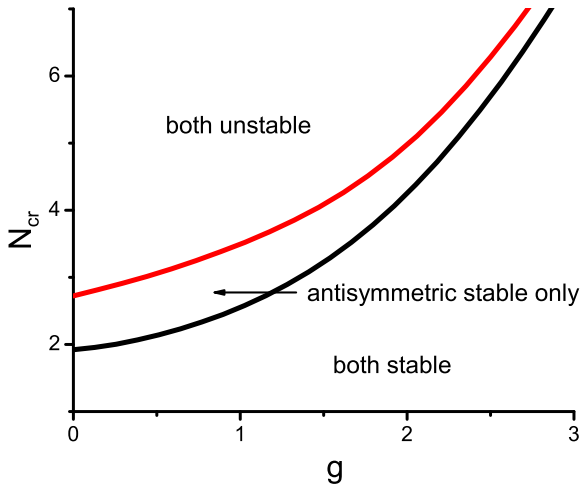}
\caption{Stability boundaries, $N_{\mathrm{cr}}(g)$, for the fundamental
(black) and first antisymmetric (red) modes trapped in the HO potential.}
\label{stability for harmonic}
\end{figure}

Results concerning stability of the fundamental and
antisymmetric states, obtained numerically, are collected in the diagram displayed in Fig. \ref%
{stability for harmonic}, in the plane of the norm and SDF cubic
coefficient, $g$. Note that the stability area expands with
the increase of $g$. Indeed, in the limit of the dominant SDF nonlinearity,
there is no apparent reason for destabilization of the fundamental and
excited states, which are obviously stable in the linear system.
Furthermore, the strengths of the destabilizing SF and stabilizing SDF terms
in Eq. (\ref{NLSE}), for modes with width $a_{\perp }\sim \Omega ^{-1/4}$
and large squared amplitude $A^{2}\sim N/a_{\perp }$, balance each other at $%
N\sim g$, which explains the asymptotically linear shape of the stability
boundaries at large $N$ in Fig. \ref{stability for harmonic}. On the other
hand we observe that the stability intervals for both the ground
state and the antisymmetric mode do not vanish in the case the SF
quintic-only interaction, $g=0$.

\section{Analytical solutions for the delta-functional trapping potential}

\label{sec:delta}

If the HO trapping potential in Eq. (\ref{U}) is replaced by the
delta-functional potential (\ref{delta}), then the respective stationary
equation (\ref{stationary NLSE}) is replaced by%
\begin{equation}
-kU+\frac{1}{2}U^{\prime \prime }-gU^{3}+U^{5}=-\epsilon \delta (x)U
\label{stat-eps}
\end{equation}%
(recall we have fixed $\epsilon =0.5$ by means of rescaling), which implies
that the jump condition (\ref{jump}) for symmetric solutions, with $%
U(-x)=U(x)$, reduces to%
\begin{equation}
U^{\prime }(x=+0)=-\epsilon U(x=0).  \label{jump-eps}
\end{equation}%
Using exact solution (\ref{solution}) of Eq. (\ref{stat-eps}) in the free
space (at $x\neq 0$), one can easily construct the following exact solution
for the mode pinned to the delta-functional potential:

\begin{widetext}
\begin{equation}
U(x,k)=\sqrt{\frac{2\sqrt{3}k}{\sqrt{4k+3g^{2}/4}\cosh {[2\sqrt{2k}(|x|+\xi(k)
)]}-\left( \sqrt{3}/2\right) g}},  \label{delta solution}
\end{equation}%
\end{widetext}
where $\xi (k)>0$ is determined by the boundary
condition (\ref{jump-eps}):%
\begin{widetext}
\begin{gather}
\xi (k)=\frac{1}{2\sqrt{2k}}\arcsin \left[ -\frac{\sqrt{3\left[ 8k^{2}+({3/2}%
)g^{2}k\right] }g\epsilon }{\left[ 8k+({3/2})g^{2}\right] (2k-\epsilon ^{2})}%
\right.  \notag \\
\left. \pm \frac{\sqrt{3\left[ 8k^{2}+({3/2})g^{2}k\right] g^{2}\epsilon
^{2}+16k\epsilon ^{2}\left[ 8k^{2}+({3/2})g^{2}k-4k\epsilon ^{2}-({3/4}%
)g^{2}\epsilon ^{2}\right] }}{\left[ 8k+({3/2})g^{2}\right] (2k-\epsilon
^{2})}\right] .  \label{xi}
\end{gather}
\end{widetext}

While at $k>0$ solution (\ref{delta solution}) is exponentially localized,
in the limit of $k=0$ remains relevant, and it produces a pair of weakly
localized pinned modes:%
\begin{eqnarray}
U(x,k &=&0)=\frac{1}{\sqrt{g\left( |x|+\xi \right) ^{2}+\frac{2}{3g}}},
\notag \\
\xi  &=&\frac{1}{2\epsilon }\pm \sqrt{\frac{1}{4\epsilon ^{2}}-\frac{2}{%
3g^{2}}}.  \label{k=0,general}
\end{eqnarray}%
Obviously, this pair exists if $g$ is large enough:%
\begin{equation}
g\geq g_{\mathrm{cr}}=2\sqrt{2/3}\epsilon \approx 0.816,  \label{cr}
\end{equation}%
where the value $\epsilon \equiv 0.5$ adopted above is substituted. The
weakly localized modes are meaningful ones as their norm converges, see
below. Note that the weakly localized state (\ref{k=0,general}) with $\xi =0$
is a valid solution to Eq. (\ref{stat-eps}) in the free space, i.e., with $%
\epsilon =0$, but in the latter case this solution is unstable \cite{Kivshar}%
.

\begin{widetext}

\begin{figure}[th]
\centering
\subfigure[] {\includegraphics[scale=0.18]{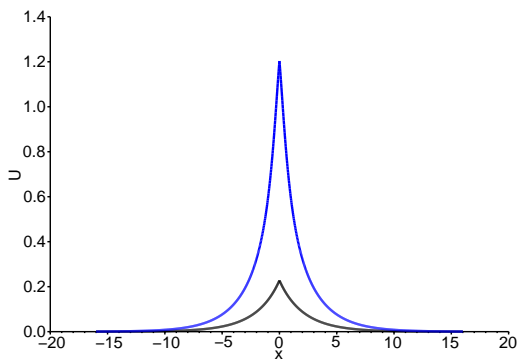}} \subfigure[] {%
\includegraphics[scale=0.18]{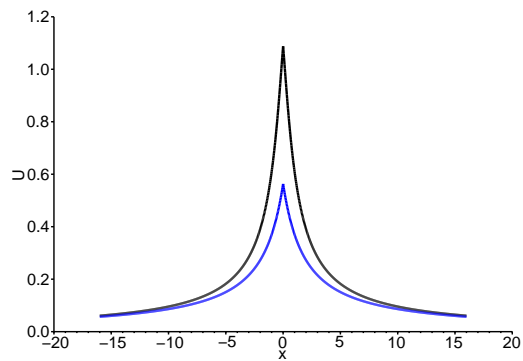}}
\caption{Exact stable solutions for the delta-functional potential (\protect
\ref{delta}), with $\protect\epsilon =0.5$, (a) given by Eqs. (\protect\ref%
{delta solution}) and (\protect\ref{xi}) with negative (black) and positive
(blue) signs for $g=1,k=0.1$, as well as (b) given by Eq. (\protect\ref%
{k=0,general}) for $g=1,k=0$ with both signs (black and blue  profiles
correspond to the solutions with \textquotedblleft -" and \textquotedblleft
+", respectively).\ \ \ \ \ \ \ \ \ \ \ \ \ \ \ \ \ \ \ \ \ \ \ \ \ \ \ \ \
\ \ \ \ \ \ \ \ \ \ \ \ \ \ \ \ \ \ \ \ \ \ \ \ \ \ \ \ \ \ \ \ \ \ \ \ \ \
\ \ \ \ \ \ \ \ \ \ \ \ \ \ \ \ \ \ \ \ \ \ \ \ \ \ \ \ \ \ \ \ \ \ \ \ \ \
\ \ \ \ \ \ \ \ \ \ \ \ \ \ \ \ \ \ \ \ \ \ \ \ \ \ \ \ \ \ \ \ \ \ \ \ \ \
\ \ \ \ \ \ \ \ \ \ \ \ \ \ .}
\label{Dirac delta solution}
\end{figure}

\end{widetext}

Examples of generic solutions (\ref{delta solution}), and the weakly
localized ones (\ref{k=0,general}), featuring a cusp at $x=0$, are displayed
in Figs. \ref{Dirac delta solution}(a) and \ref{Dirac delta solution}(b),
respectively.

At $k>0$, the analytical solution given by Eqs. (\ref{delta solution}) and (%
\ref{xi}) seems quite complex. It takes substantially  simpler form for the
particular case of the quintic-only nonlinearity, $g=0$:%
\begin{eqnarray}
U_{g=0}(x,k) &=&\left( 3k\right) ^{1/4}\sqrt{\mathrm{sech}\left( 2\sqrt{2k}%
\left( |x|+\xi \right) \right) },  \notag \\
\xi  &=&\frac{1}{2\sqrt{2k}}\mathrm{Artanh}\left( {\frac{\epsilon }{\sqrt{2k}%
}}\right) .  \label{g=0}
\end{eqnarray}%
Obviously, this solution exists for sufficiently large propagation
constants,
\begin{equation}
k>k_{\min }\equiv \epsilon ^{2}/2.  \label{kmin}
\end{equation}

In the opposite limit of the dominating SDF\ cubic term with large $g$, when
the quintic term may be omitted in Eq. (\ref{stat-eps}), the general
solution given by Eqs. (\ref{delta solution}) and (\ref{xi}) also simplifies:%
\begin{eqnarray}
U_{g\rightarrow \infty }(x) &=&\sqrt{\frac{2k}{g}}\frac{1}{\sinh \left(
\sqrt{2k}\left( |x|+\xi \right) \right) },  \notag \\
\xi  &=&\frac{1}{\sqrt{2k}}\mathrm{Artanh}\left( {\frac{\sqrt{2k}}{\epsilon }%
}\right).  \label{large-g}
\end{eqnarray}%
Its existence region is complementary to that given by Eq. (\ref{kmin}):
\begin{equation}
k\leq k_{\max }\equiv \epsilon ^{2}/2.  \label{epsilon}
\end{equation}
Solution (\ref{large-g}) with zero propagation constant, $k=0$, corresponds
to a weakly localized mode,%
\begin{equation}
U_{g\rightarrow \infty }(x,k=0)=\frac{1}{\sqrt{g}\left( |x|+\epsilon
^{-1}\right) },  \label{k=0}
\end{equation}%
which is the limit case (for large $g$) of the above solution (%
\ref{k=0,general}).

The norm of the general solution, given by Eqs. (\ref{delta solution}) and (%
\ref{xi}), is equal to 
\begin{widetext}
\begin{equation}
N(k)=\frac{2\sqrt{2k}B}{g}\left\{ \frac{1}{\sqrt{1-B^{2}}}\left[ \arcsin
B+\arcsin \left( \frac{1-B\cosh (2\sqrt{2k}\xi )}{\cosh (2\sqrt{2k}\xi )-B}%
\right) \right] \right\},  \label{general norm}
\end{equation}%
\end{widetext}
where $B\equiv g\sqrt{3/\left( 16k+3g^{2}\right) }$. In the limit of $k=0$
[provided that $g$ exceeds the critical value (\ref{cr})], the norm of the
weakly localized solution (\ref{k=0,general}) can be found in a more
explicit form,%
\begin{equation}
N(k=0)=\sqrt{6}\left[ \frac{\pi }{2}-\arctan \left( \sqrt{\frac{3}{2}}g\xi
\right) \right] .  \label{N(k=0)}
\end{equation}
Furthermore, in the case of the quintic-only nonlinearity, $g=0$, when the
solution amounts to Eq. (\ref{g=0}), expression (\ref{general norm}) reduces
to
\begin{equation}
N_{g=0}(k)=\sqrt{6}\left\{ \frac{\pi }{2}-\arctan \left[ \exp \left( \mathrm{%
Artanh}\left( {\frac{\epsilon }{\sqrt{2k}}}\right) \right) \right] \right\} .
\label{special norm}
\end{equation}%
In the other above-mentioned limit, of large $g$, when the solution
takes the form of Eq. (\ref{large-g}), the respective simplification of
expression (\ref{general norm}) for the norm reads%
\begin{equation}
N_{g\rightarrow \infty }(k)=\frac{2}{g}\left( \epsilon -\sqrt{2k}\right) .
\label{N-large-g}
\end{equation}%
Note that Eq. (\ref{N-large-g}) at $k=0$ matches Eqs. (\ref{N(k=0)}) and (%
\ref{k=0,general}) in the limit of large $g$.

\begin{widetext}

\begin{figure}[th]
\centering
\subfigure[] {\includegraphics[scale=0.32]{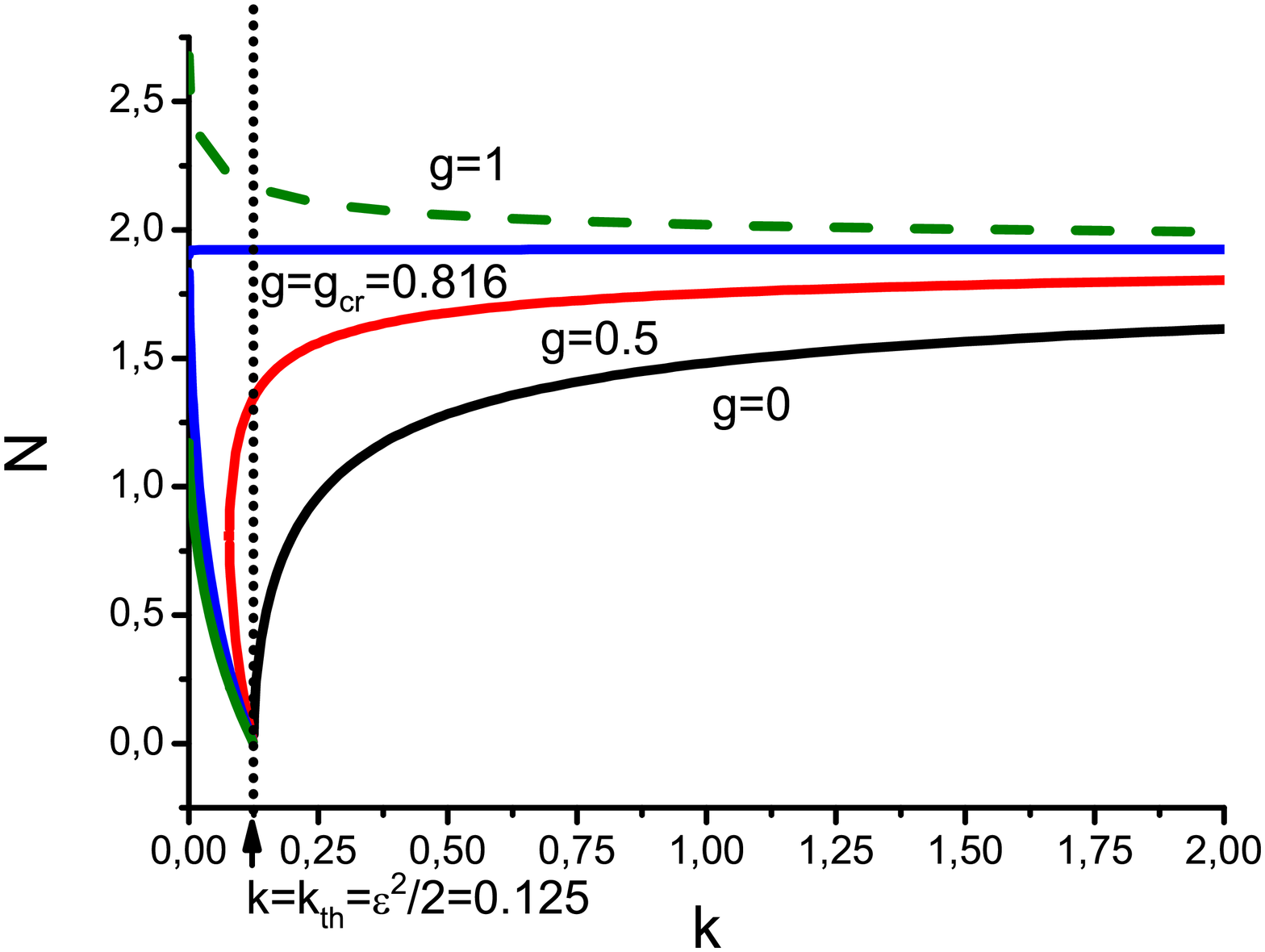}} \nolinebreak 
\subfigure[] {\includegraphics[scale=0.32]{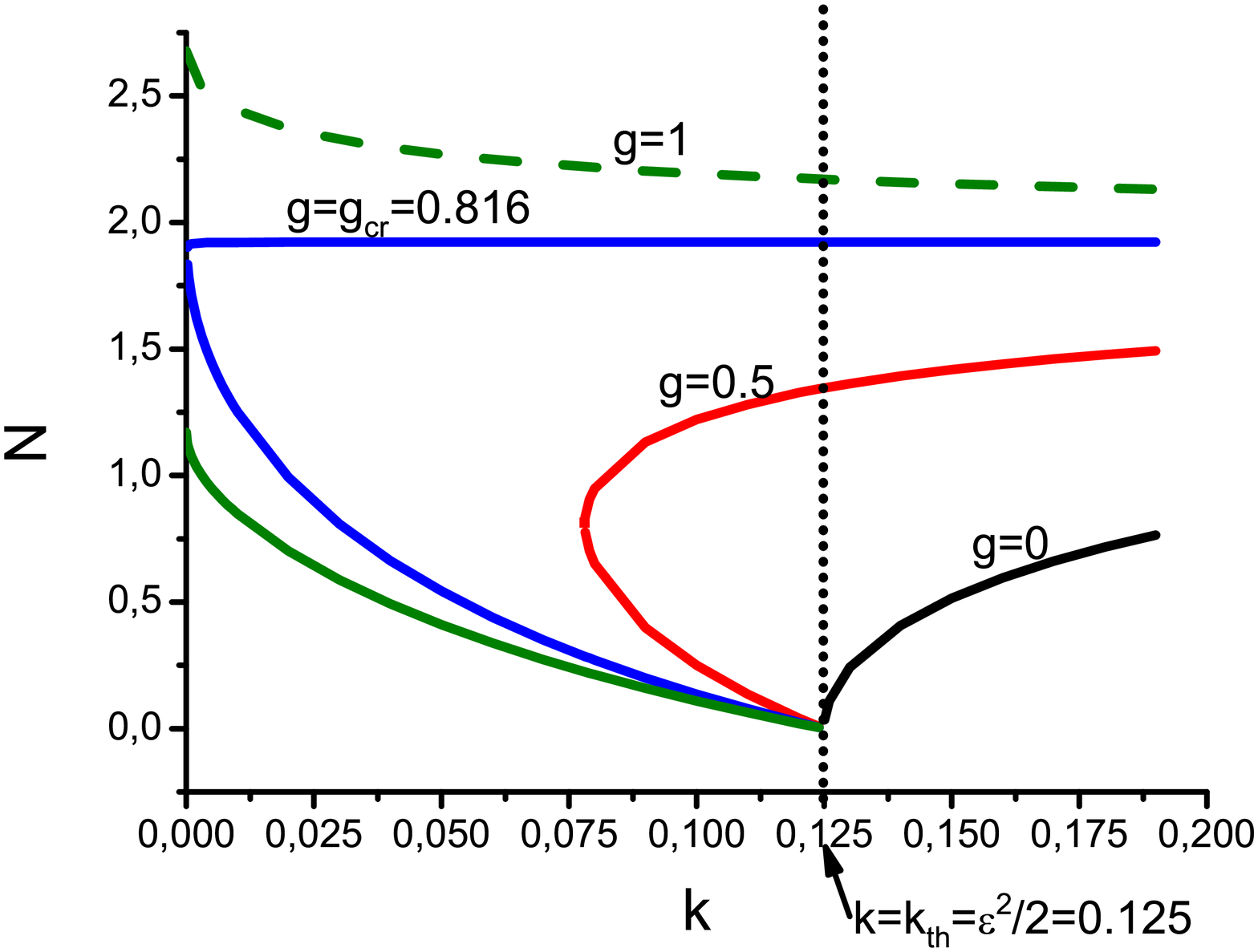}} 
\subfigure[] {\includegraphics[scale=0.32]{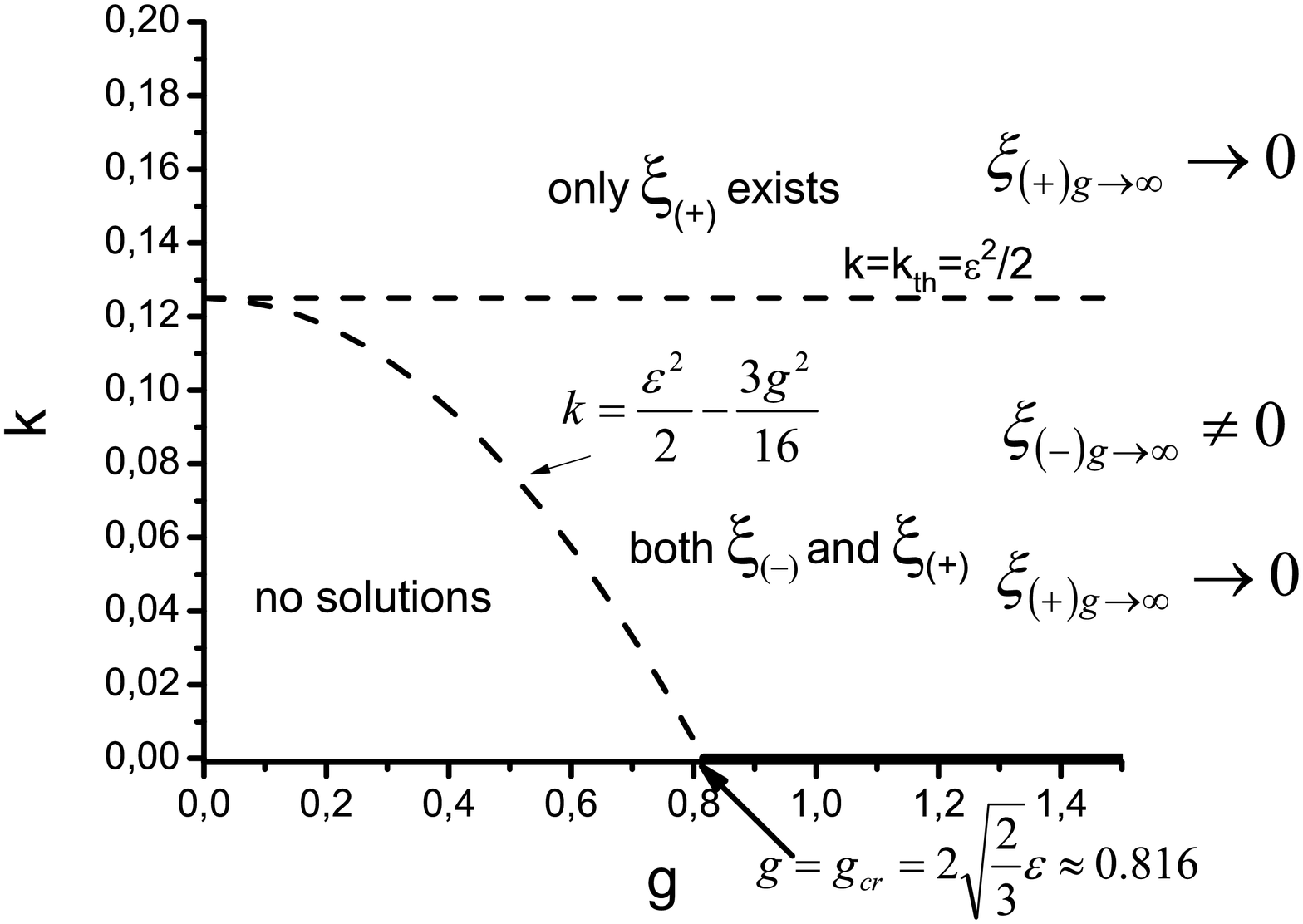}} \nolinebreak 
\subfigure[] {\includegraphics[scale=0.32]{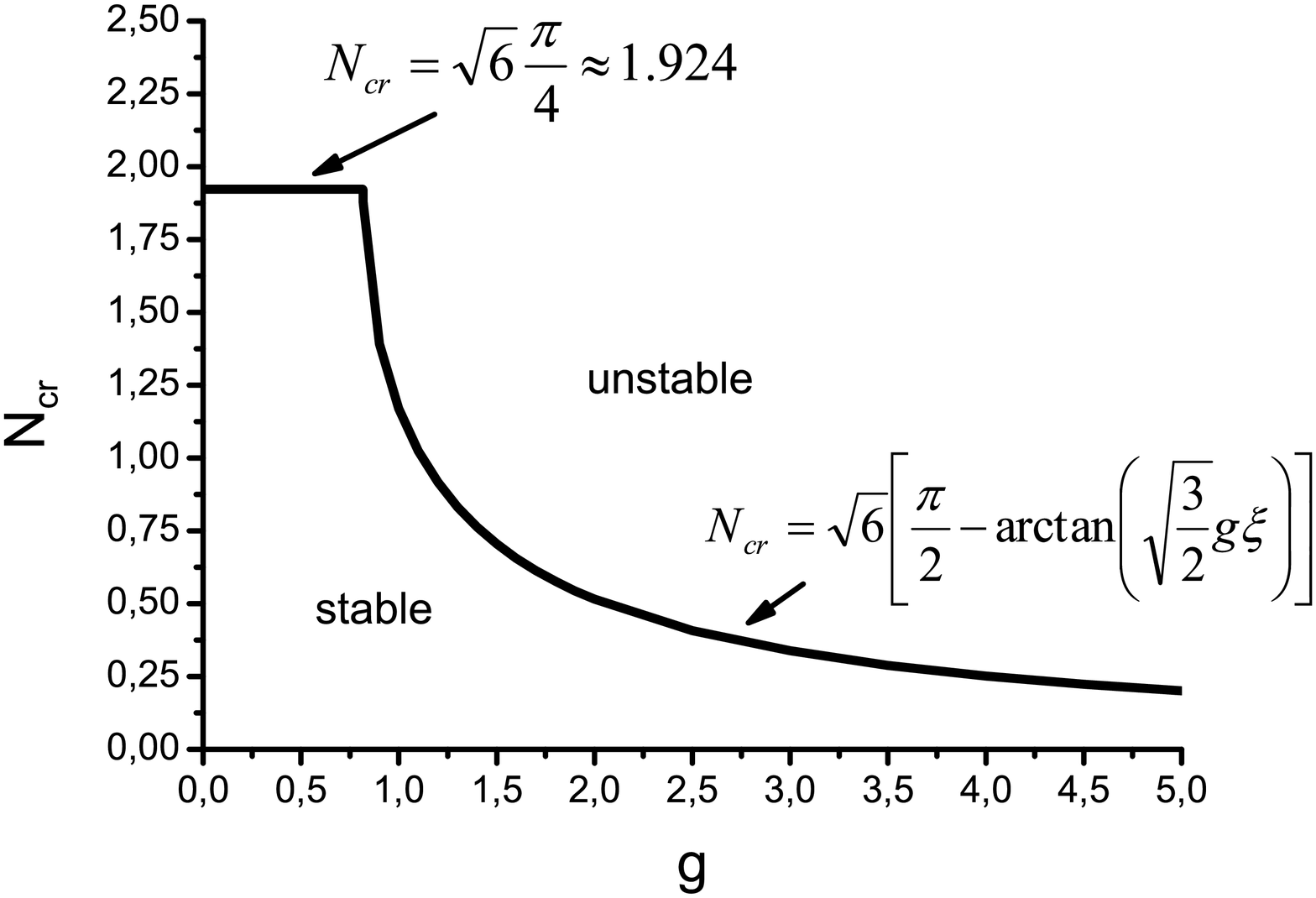}}
\caption{(a) The norm of the exact solution trapped in the delta-functional
potential (\protect\ref{delta}) vs. the propagation constant for different
values of strength $g$ of the SDF cubic term. Unstable branches are shown by
dashed lines. (b) The same as (a) in a smaller range of $k$. (c) The
existence area for the solutions in the $\left( k,g\right) $ plane. (d) The
stability boundary, shown as the critical norm vs. the SDF strength, $g$.}
\label{N(k) comparison}
\end{figure}

\end{widetext}

To check the stability of these solutions, the VK criterion can be applied
first. For the case of the purely SF (quintic-only) nonlinearity, $g=0$,
when this criterion is definitely relevant, Eq. (\ref{special norm}) yields
\begin{equation}
\frac{dN}{dk}=\frac{\sqrt{3}\epsilon }{4\sqrt{1-(\epsilon ^{2}/2k)}}k^{-3/2},
\label{norm derivative}
\end{equation}%
which is always positive in its existence range (\ref{kmin}), hence the
corresponding family of solution is entirely stable. In the general case, with $%
g>0$, the $N(k)$ dependence is plotted in Fig. \ref{N(k) comparison}(a)
and \ref{N(k) comparison}(b) (in a smaller range). 

We notice in Figs. \ref{N(k) comparison}(a)
and \ref{N(k) comparison}(b) that the branches produced by $``+"$ sign in Eqs. (\ref{delta solution}) and (\ref{xi}%
),which are located below the blue line corresponding to the critical value $%
g=g_{\mathrm{cr}}$ given by Eq. (\ref{cr}), entirely satisfy the VK
criterion, while at $g>$ $g_{\mathrm{cr}}$ they are entirely VK-unstable,
featuring $dN/dk<0$. In all \textit{forward-going} branches
predicted by Eqs. (\ref{delta solution}) and (\ref{xi}) with the $``+"$
sign, point $k=k_{\mathrm{th}}\equiv \epsilon ^{2}/2=0.125$ is a singular
one. On the other hand, \textit{backward-going} branches
corresponding to $``-"$ in Eq. (\ref{xi}) do not satisfy the VK criterion,
hence they are either completely unstable or satisfy the anti-VK\textit{\ }%
criterion, similar to the HO case. Their stability has been confirmed by the
numerical verification, hence for them the anti-VK criterion indeed
guarantees the stability. All the existing solutions (stable and unstable)
are plotted in Fig. \ref{N(k) comparison}(c), including the special ones
predicted by Eqs. (\ref{k=0,general}) for $k\rightarrow 0$. Below $k=k_{%
\mathrm{th}}$, the existence area is limited by
\begin{equation}
k=\frac{\epsilon ^{2}}{2}-\frac{3g^{2}}{16}  \label{asymptote}
\end{equation}%
which can be derived from Eq. (\ref{xi}) as the boundary of physical
solutions. Above $k_{\mathrm{th}}$, only the solution given by Eq. (\ref{xi}%
) with the positive sign exists, its negative counterpart being unphysical.
In the limit of $g\rightarrow \infty $, the solutions with $``+"$ sign in
Eq. (\ref{xi}) become irrelevant, as they require condition $\xi
_{(+)g\rightarrow \infty }\rightarrow 0$, which can be satisfied only in the
system without the delta-functional potential. The only physical solution in
this regime is the one expressed by Eqs. (\ref{delta solution}) and (\ref{xi}%
) in the case of the $``-"$ sign. In other words, in this limit there is
only one physical solution which can be obtained from Eq. (\ref{large-g}),
but, as said above, the solution with the negative sign can exist only for $%
k<k_{\mathrm{th}}$, in agreement with condition (\ref{epsilon}).

\begin{widetext}

\begin{figure}[t]
\centering\centering\subfigure[]{%
\includegraphics[height=2.5in]{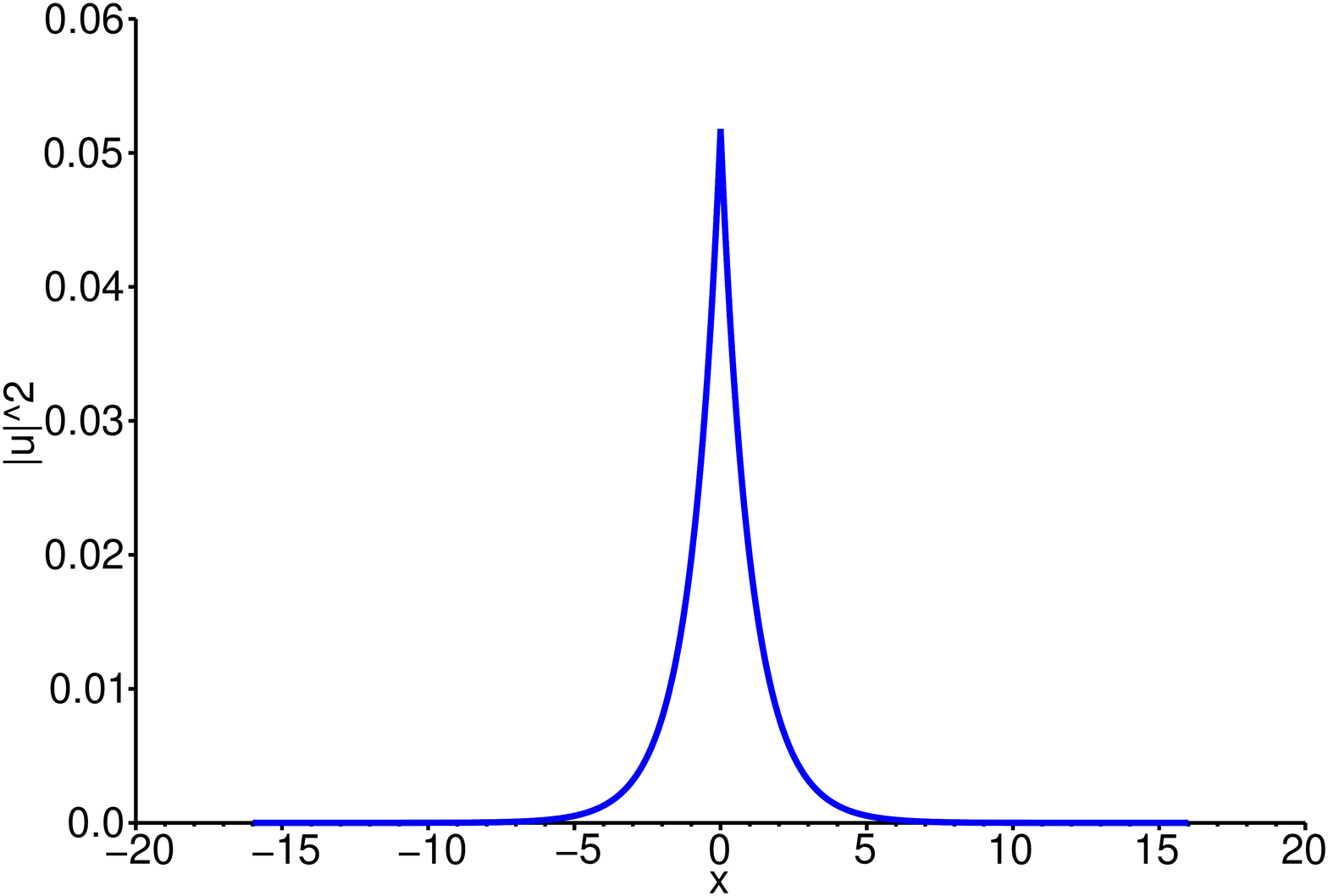}} \nolinebreak \centering
\subfigure[]{\includegraphics[height=2.5in]{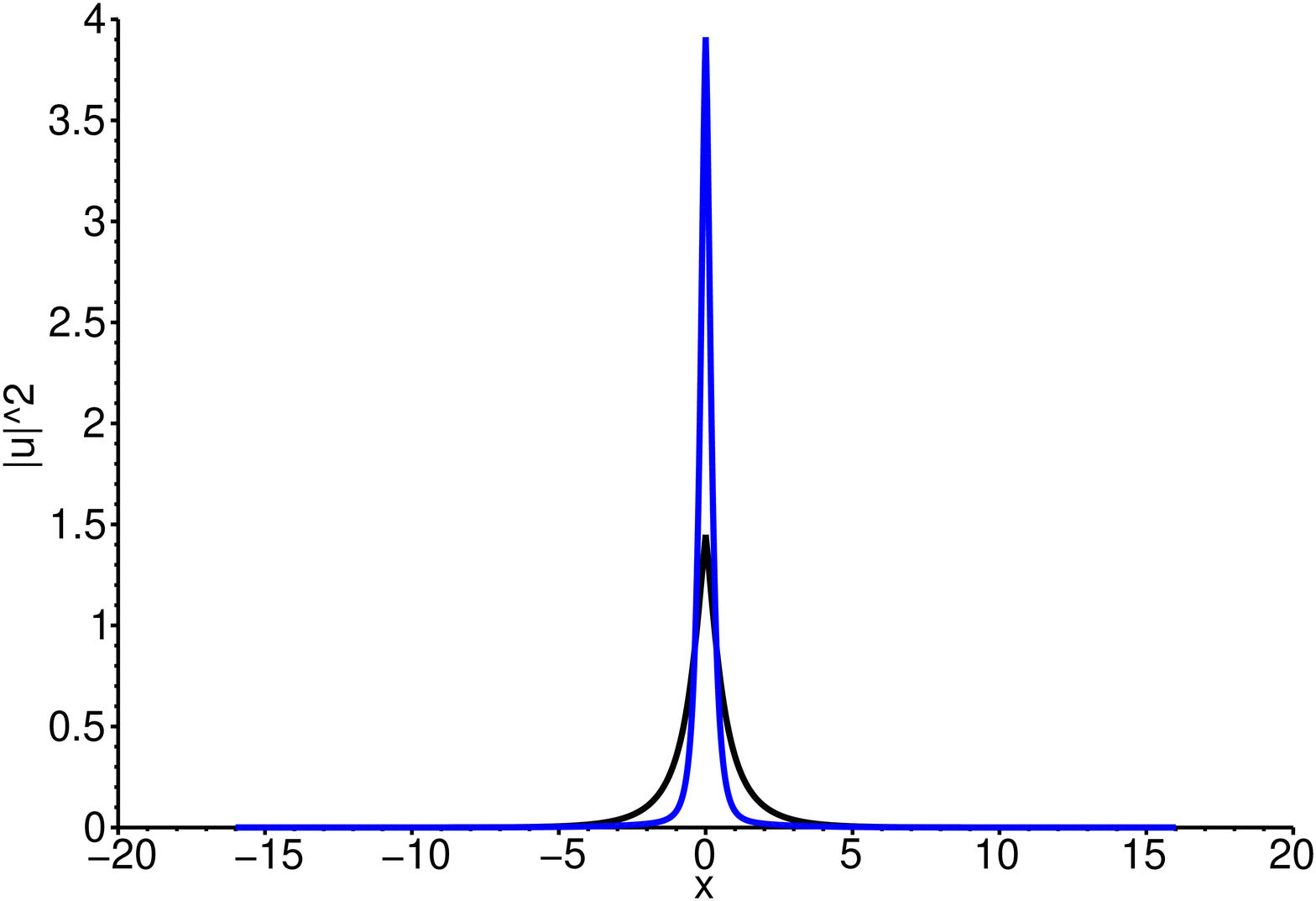}}
\caption{Examples of the perturbed evolution of stable (a) and unstable (b)
modes pinned to the delta-functional potential (\protect\ref{delta}), for $%
g=1,k=0.1,N=0.109$ and $g=1,k=0.1,N=2.193$, respectively (the same $k$, but
the solution belongs to the different branches, with the $"-"$ and $"+"$
signs). The initial and final configurations are shown by the black and blue
curves, respectively.}
\label{antisymmetric-evolution}
\end{figure}

\end{widetext}

The predictions of the VK criterion for the present setting were confirmed
in the case of the branches corresponding to the positive sign by numerical
evaluation of eigenvalues for small perturbations, approximating, in the
framework of the numerical scheme, the ideal delta-function by the
Kronecker's delta, subject to the same normalization as the delta-function.
This approximation, used for the numerical evaluation of perturbation
eigenvalues, as well as for direct simulations, gives very reliable results.
For the stable branches, e.g., at $g=0$, all the eigenfrequencies are real,
while in the case of an instability, such as the \textit{forward-going }%
branch with the positive sign at $g\geq g_{\mathrm{cr}}$, e.g., at $g=1$,
there are pure-imaginary ones. The imaginary part of the perturbations
eigenvalues vanishes for \textit{backward-going} branches, corresponding to
the negative sign, both for $g\geq g_{\mathrm{cr}}$ and $g<g_{\mathrm{cr}}$.
That conclusion confirms the anti-VK criterion for the latter branches, as
they obey condition $dN/dk<0$. Furthermore, all \textit{forward-going}
branches below $g_{\mathrm{cr}}$ asymptotically tend to the same limit which
can be easily calculated by taking the $U_{g=0}$ solution from Eq. (\ref{g=0}%
) in the limit of $k\rightarrow \infty $ and its norm given by Eq. (\ref{special norm}), $N_{g=0,k\rightarrow \infty}=\sqrt{6}\left(
\pi /4\right) \approx 1.924$.
This limit value is a stability boarder for all the branches with the
positive and negative signs at $g<g_{\mathrm{cr}}\approx 0.816$. At $g\geq g_{\mathrm{cr}}$ the stability boundary can be
found from Eq. (\ref{N(k=0)}) with $\xi _{(-)}$ taken as per Eq. (\ref%
{k=0,general}). For given $g$ this is a stable solution with the largest
possible value of norm (the largest-norm stable solution belongs to the \textit{%
backward-going} branch). Obviously, there exist other solutions with still
larger norms for the same $g$, but they belong to the \textit{forward-going}
branch, which is entirely unstable. The stability analysis presenting total
norm $N$ versus the SDF strength, $g$, is summarized in Fig. \ref{N(k)
comparison}(d).

The results were also verified by direct simulations\ of the perturbed
evolution of the modes under consideration. Typical examples of the stable
and unstable evolution are presented in Fig. \ref{antisymmetric-evolution}.
The unstable solution starts collapsing in the course of the evolution,
similar to the system with the HO trapping potential considered above. The failure of the delta-functional pinning potential to stabilize the
trapped modes at $g\rightarrow +\infty $, see Eq. (\ref{cr}), is a drastic
difference from the case of the HO potential, where the stability of the
trapped modes monotonously enhances with the increase of the strength of the
cubic SDF term, $g$, see Fig. \ref{stability for harmonic}.

Lastly, stable solutions for solitons pinned to the attractive
delta-functional potential in the model with the opposite combination of the
nonlinearities, SF cubic and SDF quintic, can also be found in an analytical
form. In particular, this setting gives rise to a bistability, with two
different pinned states corresponding to a common value of the propagation
constant \cite{GMW}.


\section{Conclusions}

The aim of this work is to study, in the analytical and numerical form, the
stabilization of 1D solitons by means of the external potential, under the
influence of the competing SF (self-focusing) quintic and SDF (self-defocusing)
cubic terms. Two standard trapping potentials were considered, the 
harmonic-oscillator one, and the Dirac's delta-function. In the former
case, both fundamental symmetric and the lowest antisymmetric modes were
studied. In the latter, the results were obtained in the completely
analytical, although somewhat cumbersome, form. The most essential result
concerns the way the competing necessary stability criteria, namely, the VK
(Vakhitov-Kolokolov) and anti-VK ones, divide the regions of their validity
in the system with the competing nonlinear terms. In particular, in both
models with the HO and delta-functional trap, the forward- and
backward-going soliton branches, in terms of their dependence $N(k)$, between
their norm and propagation constant, precisely obey the VK and anti-VK
criteria, respectively.

It is relevant to extend the analysis for more generic forms of the trapping
potential, taking into regard that the cases of the broad HO and narrow
delta-functional traps produce very different results. In the 2D geometry,
it may be interesting to consider the stabilization provided by trapping
potentials in models combining the SF cubic nonlinearity (recall it leads to
the collapse in 2D) and effectively SDF quadratic interactions. This system
may be a relevant model of an optical bulk waveguide.

\nocite{*}
\begin{acknowledgments}
B.A.M. and M.A.K. acknowledge partial support from the National Science Center of Poland in the frame of HARMONIA program no. 2012/06/M/ST2/00479. K.B.Z. acknowledges support from the National Science Center of Poland in project ETIUDA no. 2013/08/T/ST2/00627. M.T. acknowledges the support of the National Science Center grant  N202 167840.
\end{acknowledgments}
\bibliography{competing}

\end{document}